\documentclass[aps,prl,superscriptaddress,twocolumn]{revtex4-1}
\pdfoutput=1

\usepackage[bookmarks=true]{hyperref} 
\usepackage{amsmath,amsfonts,amsmath}

\usepackage{graphicx,subfigure,epsfig,color}

\usepackage[T1]{fontenc}
\usepackage{lmodern}
\usepackage{amsmath}
\usepackage{amsthm}
\usepackage{amsfonts}
\usepackage{amssymb}
\usepackage{mathrsfs}
\usepackage{physics}
\usepackage{graphicx}
\usepackage{caption}
\usepackage{enumitem}
\usepackage{framed,float}
\usepackage{booktabs}
\usepackage{array}
\usepackage{soul}
\usepackage{multirow}
\usepackage{comment}
\usepackage{braket}
\usepackage{todonotes}
\usepackage{youngtab}
\usepackage{tikz}

\def \be  {\begin{equation}}
\def \ee  {\end{equation}}
\def \ba {\begin{equation}\begin{aligned}}
\def \ea {\end{aligned}\end{equation}}
\def \bea  {\begin{eqnarray}}
\def \eea  {\end{eqnarray}}

\newcommand{\nn}{\nonumber}

\newcommand{\cN}{\mathcal{N}}

\usepackage{color}

\begin{document}

\allowdisplaybreaks

\title{Deriving Neutron Star Equation of State from AdS/QCD}

\author{Wei Li}
\email{wei_li_phys@163.com}
\affiliation{Department of Physics and Institute for Quantum Science and Technology, Shanghai University, 99 Shangda Road, Shanghai 200444, China}

\author{Jing-Yi Wu}
\email{wujingyi222@mails.ucas.ac.cn}
\affiliation{School of Astronomy and Space Science, University of 
Chinese Academy of Sciences (UCAS), Beijing 100049, China}

\author{Kilar Zhang}
\email{kilar@shu.edu.cn, Corresponding Author}
\affiliation{Department of Physics and Institute for Quantum Science and Technology, Shanghai University, 99 Shangda Road, Shanghai 200444, China}
\affiliation{Shanghai Key Lab for Astrophysics, Shanghai 200234, China}
\affiliation{Shanghai Key Laboratory of High Temperature Superconductors, Shanghai 200444, China}


\begin{abstract}
Neutron stars are among the main targets for gravitational wave observatories, however, their equation of state is still not well established. Mainly phenomenological models with many parameters are widely used by far, while theoretical models are not so practical. In arXiv:1902.08477, a theoretical equation of state with only one parameter is derived from Witten-Sakai-Sugimoto model, as an application of AdS/QCD, where pointlike instanton case is taken into consideration. When the  tidal deformability constraint from gravitational wave event is satisfied, the maximum mass is about 1.7 solar masses. Now we upgrade this model to instanton gas, with one more variable, the instanton width. This is not naively a free parameter, but a function of the chemical potential. Thus we end up with a more complicated and accurate model, but still with only one adjustable parameter. In this case, we find the maximum mass becomes 1.85 solar masses. This is an encouraging result, as a theoretically derived model. 
\end{abstract}

\maketitle

\section{I. Introduction}
The observations of neutron stars (NS) based on electromagnetic (EM) waves \cite{Margalit:2019dpi} have a long history, yet the equation of state (EoS) of NS has remained as a puzzle. The precise derivation of EoS requires much higher density than neutron saturation level, the region where no perturbative methods work and almost impossible to settle. Nonetheless, there are still many phenomenological or approximate models.   With EoS in hand, one can find the Mass-Radius relation and also the tidal Love number (TLN), by solving the Tolman-Oppenheimer-Volkoff (TOV) equations together \cite{Hinderer_2009}. That means, on the contrary, if we have the information of Mass-Radius or TLN, we can constrain the EoS, thus offering an alternative method to probe the high density region in nuclear physics \cite{Lattimer:2012nd}.

Then, what is the situation on the observation side? We know it is easy to read off the masses of NS to high accuracy, especially for (binary) pulsars \cite{2009Physics}. While for the radius, realities are much less optimistic. Only through X-ray bursts \cite{Galloway:2007dn} or thermal dynamics, could we find some limited information on the radius, with an error bar of several kilometers for a standard $1.4 M_\odot$ NS \cite{2020Sci...370.1450D}. This could barely serve as a constraint, so there are more than one hundred common EoS models: no one violates the requirements. Recent progress from the NICER mission \cite{Miller:2021qha} shows $12.45 \pm 0.65$ km for a $1.4 M_\odot$ NS, but this result is EoS dependent, and also make use of the Gravitational Waves (GW) data. 

Since the completions of GW observatories, especially after the first NS GW event, we enter the era of multi-messenger astronomy \cite{2017Multi}. The advantage of GW data is not only its high efficiency in finding black holes (BH) and NS, but also an extra parameter that beyond the normal EM observation ability: the TLN.  This TLN reflects the deformability of a compact star under an external quadratic force. In a sense, it shows the compactness of a star. Its theoretical value can actually be deduced from a second order perturbation of Einstein equation, along with the EoS. As a result, if we know the value of TLN from observation, inversely we could constrain the EoS. In fact, the TLN from GW offers very strong constraints, and even merely by the First Binary NS event GW170817 \cite{LIGOScientific:2017vwq,LIGOScientific:2018hze}. Half (including one marginal) of the 7 major EoS candidates are ruled out.
So far the story is so good, yet no further strong TLN constraints are obtained, since TLN is 5 Post-Newtonian and difficult to obtain for distant events. For another NS events GW190425 \cite{LIGOScientific:2020aai}, one have TLN much higher with only one upper limit, while for others, we don’t have any data. The window for EoS is still big enough for half of the original candidates to survive,  say some 50 kinds. 

Furthermore, people are not satisfied with pure phenomenological models, and the explore on theoretical derivations continues. Starting from the MIT bag model \cite{Rezaei:2024ydg}, many efforts have been made. In \cite{Zhang:2019tqd}, a new perspective is tried: EoS can be extracted from Witten-Sakai-Sugimoto model (WSS model) \cite{Witten:1998zw, Sakai:2005yt, Sakai:2004cn},
an application of the anti-de Sitter/quantum chromodynamics (AdS/QCD) duality\cite{Witten:1998zw, Polchinski:2000uf}, originated from superstring theory. 

AdS/QCD is an application of anti-de Sitter/conformal field theory (AdS/CFT) correspondence \cite{Maldacena:1997re}, in the sense that string theory in 5D AdS space is dual to the $\cN = 4$ super Yang–Mills theory, which is very close to QCD and share many common properties, when taking the large $N_c $ (color number) limit. The seeming surprising relation is in a sense very natural, since the initial goal of string theory is to explain strong interactions. To add the flavor parameter, \cite{Karch:2002sh} introduces flavor branes probe with negligible back reaction, which is established when the flavor number $N_f$ is much smaller than $N_c$. Though still not exact QCD, the WSS model is by now the closest model as a "top-down" approach resembling the low-energy properties of QCD, with the supergravity limit preserved.  The duality offers a way to deal with the strong coupled region in nuclear physics by the help of brane constructions. The main idea is to use D4/D8-branes, and the strings connecting different branes. The D4-branes are wrapped on a Scherk-Schwarz circle \cite{Witten:1998zw}, while the (anti-) D8-branes are added transverse to the circle as flavor probes. Then the chiral restoration can be described by the geometric separation of the D8- and $\overline{\rm D8}$- (anti-D8) branes \cite{ Sakai:2005yt, Sakai:2004cn, Bergman:2007wp, Bergman:2008qv}.

In \cite{Zhang:2019tqd} a simplified case is taken into consideration: the pointlike instanton case \cite{Li_2015, Bergman:2007wp}. The EoS derived has only one adjusting parameter: the maximum separation distance between the branes. This is a great advantage compared to other models with numerous coefficients, that you do not induce any parameter by hand. By generating the M-R relation and $\Lambda$-$\rm{M}$ relations, it turns out that if the TLN constraint is satisfied, the maximal mass supported by this EoS is $1.7 M_\odot$ (solar masses), unfortunately not enough compared with the known observation of NS with over $2 M_\odot$. 

Certainly, there are some more realistic models used to extract neutron EoS, like SLy4 \cite{Douchin:2001sv}, APR4 \cite{Akmal:1998cf} etc., but with many adjustable parameters, and need hypotheses and approximations for dense region. We want a top-down approach based on a rigorous string theory, and the holography method is ideal for the strong coupling calculation. Besides, our model has only one parameter (actually there are two more, the 't Hooft coupling $\lambda_{Y M}$ and the Kaluza-Klein mass $M_{kk}$ \cite{Duff:1986hr}, which we just choose their typical value).

Among the holographic approach there are also many interesting studies. D3/D7 model is discussed in\cite{Hoyos:2016zke,Hoyos:2016cob,Annala:2017tqz}, which needs additional inputs to break the conformal sound barrier. 
Besides, a seminal work \cite{Aleixo:2023lue} on D3/D7 model explores the static and dynamical properties of quark star, demonstrating that the LIGO-Virgo and NICER constraints cannot be satisfied simultaneously in this case.
Other efforts mainly focus on the homogeneous ansatz case, in the Hard-Wall model \cite{Bartolini:2022rkl} or in  the WSS model \cite{Bartolini:2023wis, Bartolini:2022gdf}, leading to realistic hybrid stars accounting also for the abundances of protons/neutrons and leptons. Another relevant work applying WSS model is the continuation of \cite{Li_2015} in \cite{Kovensky:2021kzl}, including the proton and lepton fractions as well as a crust modelled with a phenomenological surface tension. Those approaches in general need constructing hybrid stars or mixed stars, so it is worth considering to what extend a single component EoS can go.  Besides, the individual instantons considered in our method is complementary to the  homogeneous ansatz case studied in the above works.

With these motivations, in this paper, we further upgrade the derivation in \cite{Zhang:2019tqd} to the more complicated instanton gas case \cite{Li_2015, Ghoroku:2012am}, with one more parameter, the instanton width, which was fixed to zero in the pointlike instanton case. 
This introduces much more difficulties in calculation, since the once single variation now becomes a triple-variation, which needs more effort to solve. 

A naive guess would be that this new parameter helps to enlarge the parameter space, thus the maximum mass could be lifted by combining the two parameters properly. However, it turns out that the instanton width is not a free parameter, but a function of the chemical potential. Thus, for instanton gas case, though the model is more accurate and complicated, we still end up with only one adjusting parameter, i.e. the brane separation distance. Then, by similar process, we generate the M-R and $\Lambda$-$\rm{M}$ relation by applying this EoS, and find that when satisfying the TLN constraints, the maximum mass is $1.85 M_\odot$. This is higher than the pointlike case discussed before, yet still below the criteria of $2 M_\odot$ from NS observation. 

Nonetheless, this fact does cannot deny the power of this EoS. Considering that a real neutron star contains more than pure neutron, a hybrid (multi layer)  or mixed model \cite{Zhang:2020dfi,Zhang:2020pfh} with different compositions should be more realistic, which could circumference the maximum mass problem.

View differently, although very inspiring, WSS model still cannot grab the whole picture of NS EoS, as there are some minor flaws about this nice theory from the beginning, that need to  think over more seriously. First,this duality is supposed to be exact only when  $N_c$ goes to infinity, yet in practice we say $N_c=3$ is large enough; in this instanton gas  model we have second order phase transition, rather than the first order transition (the homogeneous ansatz case does reproduce first order transition, but has no chiral restoration), which can be seen as the reminiscent of Bose-Einstein condensation \cite{Preis:2011sp}. 

Other Known limitations of the current model are shown in \cite{Li_2015}: we are using the classical gravity approximation which works for large $\lambda_{Y M}$, while it is the small  $\lambda_{Y M}$ limit dual to large-$N_c$ QCD. The dual field theory for the deconfined geometry we apply in this paper is not clear, yet the confined geometry is connected to the known confined phase of the dual theory. The flavor brane back-reaction on the background geometry is ignored, since we use the probe brane limit. Besides, the prescription for the non-Abelian Dirac-Born-Infeld action is not uniquely defined. If we want a more realistic model, we need improve the WSS model itself, or construct a new holographic model from the beginning. We leave this to the future work.

As an alternate option, we may turn to other approaches \cite{Cai:2022omk, Zhao:2023gur} consistent with lattice QCD, or construct NS with holographic multiquarks core\cite{Pinkanjanarod:2020mgi,Burikham:2021xpn}.

The organization of the paper is as the following: After this introduction in section 1, we briefly review the main idea of WSS model and introduce our constructions for EoS extraction in section 2. In section 3 the M-R and $\Lambda$-$\rm{M}$ relations are shown, and we apply the GW constraints. We conclude in section 4. Some calculation details are collected in the appendix.

\section{II. Extracting Neutron Star Equation of State from Witten-Sakai-Sugimoto Model}
\subsection{II.1 Witten-Sakai-Sugimoto model}

Our construction is based on the "top-down" model of holographic quantum chromodynamics originated from string theory: WSS model. Following the procedure in \cite{Li_2015}, we collect the necessary basics. 

We consider $N_f$ D8- and $N_f$ $\overline{\rm D8}$-branes around $N_c$ D4-branes background which connect at a tip $u=u_c$ and maximally separated at $u=\infty$, with $u$ the (dimensionless) holographic coordinate. The asymptotic separation $\ell= \frac{\pi}{M_{kk}}$ for confined geometry, where $M_{kk}$ is the Kaluza-Klein mass. For deconfined case, however, $\ell$ is a free parameter. In this paper, we deal with the deconfined case while the confined model can be obtained as a special case.

$x_4$ represents the coordinate of the compactified dimension, where the radius is $1\over M_{kk}$. Thus, its periodicity is given as $x_4= x_4 +{2\pi\over M_{kk}}$. Notice that we are actually using the dimensionless counter parts as lower case terms, and all the equations should be understood as dimensional form in \cite{Li_2015}, like $X_4= X_4 +{2\pi\over M_{kk}}$. 
This separation boundary condition can be mathematically described as 
\ba
\ell=2 \int_{u_c}^\infty du x_4'\ ,
\ea
where the derivative mark $'$ is taken with respect to $u$.

The Abelian part of the gauge fields are denoted by  $\hat{a}_\mu$. The chemical potential $\mu$ is obtained by the 0 component of $\hat{a}_\mu$ with $\mu=\hat{a}_0 ({\infty})$.

Considering the gauge field action  consisting a Dirac-Born-Infeld (DBI) and a Chern-Simons (CS) contribution,
$
S=S_{DBI}+S_{CS}
$,
the free energy density $\Omega $ can be written as 
\ba 
\frac{\Omega}{\cal N} 
= \int_{u_c}^\infty du\,u^{5/2}\zeta\left[1+g_2+\frac{(n_IQ)^2}{u^5}\right] - \mu n_I \, ,
\ea
with the abbreviation
\ba \label{defzeta0}
\zeta \equiv \sqrt{\frac{1+g_1+u^3x_4'^2-\hat{a}_0'^2}{1+g_2}} = \frac{u^{11/2}x_4'}{k}\\ = \frac{u^{5/2}\hat{a}_0'}{n_IQ} 
= \frac{\sqrt{1+g_1}}{\sqrt{1+g_2-\frac{k^2}{u^8}+\frac{(n_IQ)^2}{u^5}}} \, . 
\ea
Here $n_I$ is the baryon number density. $\textit{k}$ is an auxiliary integration constant introduced during taking the equations of motion for $\hat{a}_0$. The detailed definition of other symbols $g_1$, $g_2$ and $Q$ are listed in appendix A.

In comparison with the pointlike instanton case considered in \cite{Zhang:2019tqd}, we now upgrade the method to instanton gas case, with an extra parameter, the instanton width $\rho$. This makes the model more realistic, though also more complicated to solve.

The temperature is set to zero to make life easier, since it is confirmed in \cite{Zhang:2019tqd} that at least for point-like case, a finite temperature only has minor effect: taking a considerably high temperature of $T=47.45$ MeV, all four coefficients in the double-polytropic EoS are altered within $2.3\%$.


By taking variations of the free energy, we obtain the following equations \cite{Li_2015}:
\begin{subequations} \label{mif0}
\bea
 \mu&=&  \int_{u_c}^\infty du\, \left[\frac{u^{5/2}}{2}\left(\frac{\partial g_1}{\partial n_I}\zeta^{-1}
+\frac{\partial g_2}{\partial n_I}\zeta \right) + \hat{a}_0'Q\right] \, , \label{mif1}\\[2ex]
0 &=&  \int_{u_c}^\infty du\, \left[\frac{u^{5/2}}{2}\left(\frac{\partial g_1}{\partial \rho}\zeta^{-1}
+\frac{\partial g_2}{\partial \rho}\zeta \right) + n_I \hat{a}_0'\frac{\partial Q}{\partial \rho}\right] \, ,\label{mif2} \\[2ex]
0&=&\int_{u_c}^{\infty} du\bigg[u^{5 / 2} \frac{\zeta g_2(p+\frac{2}{u_c})+\zeta^{-1} g_1(p-\frac{2}{u_c})}{2}\\
&-&n_I q \hat{a}_0^{\prime} \frac{u^3+2 u_c^3}{3 u^2 u_c}-\frac{\alpha k}{6 u_c^2 c_1(u-u_c)^{3 / 2}}+\frac{3 u_c^2}{u^{1 / 2} f_c} \frac{\zeta^{-1} g_1}{2}\bigg] \nn \, . \label{mif3}
\eea
\end{subequations}

Then by tedious numerical calculations, the variables as functions of $\mu$ can be obtained. Especially, we overcome the numerical difficulties accoutered in \cite{Li_2015} when $\mu$ is low, which is crucial in determining the low pressure behavior of the EoS. The overall frame for the numerical calculation is shown in appendix A.


After the free energy $\Omega(\mu)$ is obtained,
the pressure $p_{Q C D}$ and energy density $\epsilon_{Q C D}$ can be determined  through the standard thermodynamic relations (at zero temperature):
\ba\label{}
p_{Q C D}=-\Omega,   \quad \epsilon_{Q C D}=\Omega+n_I \mu\,.
\ea

\subsection{II.2 Equation of state}
With $p_{Q C D}(\mu)$ and $\epsilon_{Q C D}(\mu)$ in hand, we can then extract the EoS. Though the result is numerical, it is accurate enough to fit it into an analytic form. 
We choose the range $p_{Q C D} \in[0,0.05]$, since both the core pressure of NS and the baryon density are typical after recovering the dimensions, as explained in \cite{Zhang:2019tqd}. In this range, the EoS can be described by the following double-polytropic function:
\ba\label{eq6}
\epsilon_{Q C D 1}=0.140 p_{Q C D 1}^{0.429}+3.896 p_{Q C D 1}^{1.335}, \quad p_{Q C D 1} \in[0,0.05].
\ea

We then recover the dimension by noticing that
\ba\label{pres}
p=c^2 \mathcal{N} \ell^{-7} p_{Q C D}, \quad  \epsilon=\mathcal{N} \ell^{-7} \epsilon_{Q C D}\,,
\ea
with $c$ the speed of light, and $\mathcal{N}=\frac{N_c \lambda_{Y M}^3 M_{K K}^4}{12(2\pi)^5} $. 
By selecting  the typical values $\lambda_{Y M} N_c \simeq 24.9$, $M_{K K} \simeq 949 \mathrm{MeV}$, we take
$
 \mathcal{N}=1.2 \times 10^{10} \mathrm{MeV}^4.
$ Thus, the only adjustable parameter left is $\ell$.

In order to apply this EoS more conveniently to NS, we rewrite EoS in terms of the astrophysical units:
\ba
r_{\odot}=G_N M_{\odot} / c^2, \quad \epsilon_{\odot}=M_{\odot} / r_{\odot}^3, \quad p_{\odot}=c^2 \epsilon_{\odot}\,.
\ea
 Then \eqref{pres}  becomes:
\ba
p / p_{\odot}=\mathcal{A}\; p_{Q C D}, \quad \epsilon / \epsilon_{\odot}=\mathcal{A} \;\epsilon_{Q C D},
\ea
where $
\mathcal{A}=1.8 \times 10^{-5} \times \ell^{-7}
$.

Consequently, the dimensionless EoS \eqref{eq6} can be converted in the astrophysical units:
{\small
\bea
{\epsilon_1 \over \epsilon_{\odot}}=
0.140 \mathcal{A}^{0.571}\left({p_1 \over p_{\odot}}\right)^{0.429}+3.896 \mathcal{A}^{-0.335}\left({p_1 \over p_{\odot}}\right)^{1.335} .\label{eos1}
\eea
}

We also listed the pointlike result in \cite{Zhang:2019tqd}, for comparison later.

{\small
\bea
{\epsilon_2 \over \epsilon_{\odot}}=
0.131 \mathcal{A}^{0.544}\left({p_2 \over p_{\odot}}\right)^{0.456}+2.629 \mathcal{A}^{-0.192}\left({p_2 \over p_{\odot}}\right)^{1.192} .\label{eos2}
\eea
}

\section{III. Gravitation Wave Constraints}
The inner structure of NS can be solved with the help of the 
TOV equations given below, which can be obtained from perturbating the Einstein equation:
\ba
\begin{aligned}
& \frac{d m}{d r}=4 \pi r^2 \epsilon\,,
\\
& \frac{d p}{d r}=-G_N\left(\epsilon+p / c^2\right) \frac{d \phi}{d r}\,,
\\
& \frac{d \phi}{d r}=\frac{m+4 \pi r^3 p / c^2}{r\left(r-2 G_N m / c^2\right)}\,, \label{tov}
\end{aligned}
\ea
where $\phi$ is the metric potential and can be decoupled from the above. Altogether we have 4 variables as functions or $r$, but with only three equations. It is the EoS that fit the missing piece, which shows the relation between the energy density and the pressure. 

What is more, there is another important property of the star, the TLN, or tidal deformability, which
 is defined as the dimensionless coefficient $\Lambda$ introduced as
\ba
Q_{i j}=-\Lambda \, M^5 \, \mathcal{E}_{i j} , \label{tln}
\ea
with $M$ the star mass, $Q_{i j}$ the induced quadrupole moment, and $\mathcal{E}_{i j}$ the external tidal field strength. This show the deformability of a star under an external gravitational field.
By higher perturbations of Einstein equation, the TLN can also be calculated, under given EoS. More details are given in appendix B.

\begin{figure}[htpb!]
\includegraphics[width=8.6cm]{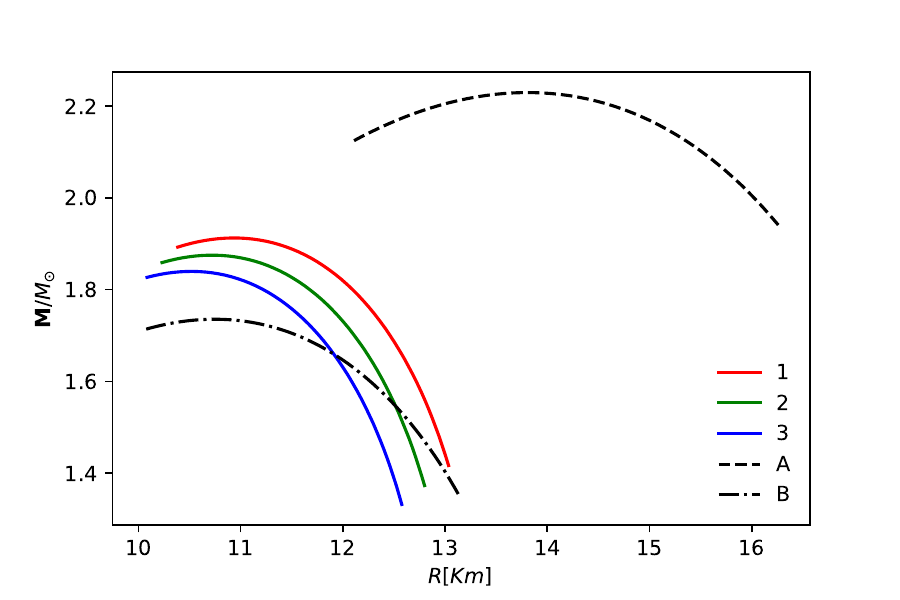}
    \caption{Mass ($\rm{M}$) vs Radius ($\rm{R}$) relations for different EoS. The color lines are obtained by EoS\eqref{eos1} with parameter $\ell^{-7}=9900,\ 10300,\ 10700$, which are labeled from 1 to 3. The results of EoS\eqref{eos2} are plotted in black lines $A$ and $B$ with $\ell^{-7}=10300,\ 17000$. }
    \label{fig1}
\end{figure}

With EoS in Eqs.(\ref{eos1}) and Eqs.(\ref{eos2}) and TOV equations (\ref{tov}), we plot FIG. \ref{fig1} by setting different values of adjustable parameter $\ell$.
FIG. \ref{fig1} shows the relationship between mass and radius. We can see five curves, labeled as 1,2,3 for instanton gas case while $\textit{A}$, $\textit{B}$ for pointlike case. For curves 1,2,3, we choose values as $\ell^{-7}=9900, 10300$, and $10700$, respectively. For $\textit{A}$ and $\textit{B}$, the values are $\ell^{-7}=10300$ and $17000$. We see that the maximal mass for instanton gas model with the three values are ranging from $1.8 M_\odot$ to $1.9 M_\odot$. In contrast, solving EoS for pointlike baryons with $\ell^{-7}=17000$, the maximal mass of NS is lower which is about $1.7 M_\odot$. It is seen from the tendency of maximal mass and $\ell$ that we can adjust the mass to exceed $2 M_\odot$ by lowering $\ell$, which is closer to the upper limit of NS mass. Actually, there exists another curve for pointlike case whose maximal mass is up to $2.2 M_\odot$. In principle any mass can be achieved by adjusting $\ell$, since there exists an scaling symmetry in M-R and $\Lambda$-M relations, as shown in \cite{Maselli:2017vfi, Zhang:2023hxd}.

The reason we did not utilize EoS of more massive NS is to limit the radius of $1.4 M_\odot$ NS to below 13km, as confirmed by multi-messenger observations \cite{2017Multi}. On the other hand, we choose such special values of $\ell$ so that the tidal deformability can meet the constraint from the analysis of GW170817. 
There are also other GW NS event like GW190425, but the tidal deformability window are much bigger, so we only need to apply the one from GW170817.  

\begin{figure}[htbp!]
\includegraphics[width=8.6cm]{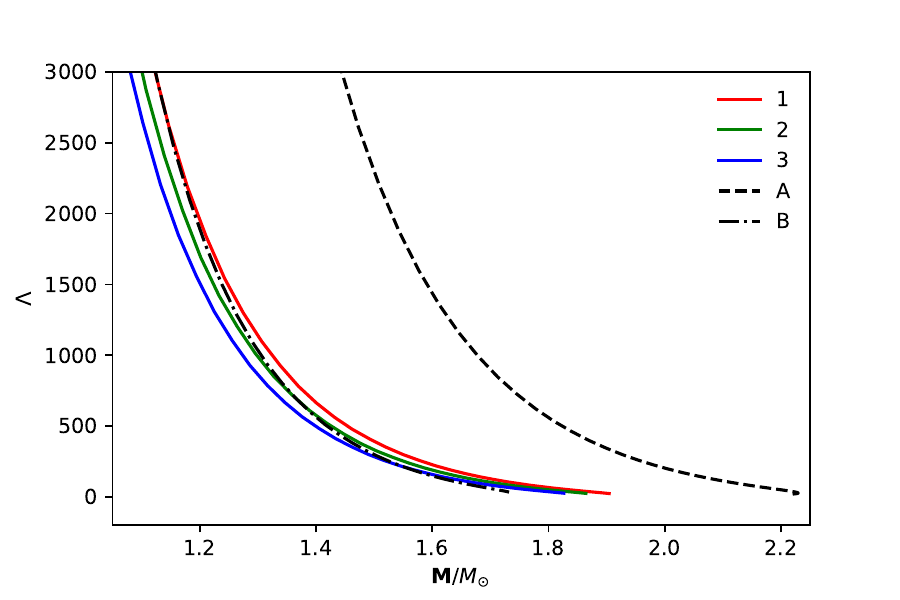}
    \caption{Tidal deformability ($\Lambda$) vs Mass ($\rm{M}$) for different EoS
with the same sets of values and labels for $\ell^{-7}$ as in FIG. \ref{fig1}.}
    \label{fig2}
\end{figure}

\begin{figure}[htbp!]
    \includegraphics[width=8.6cm]{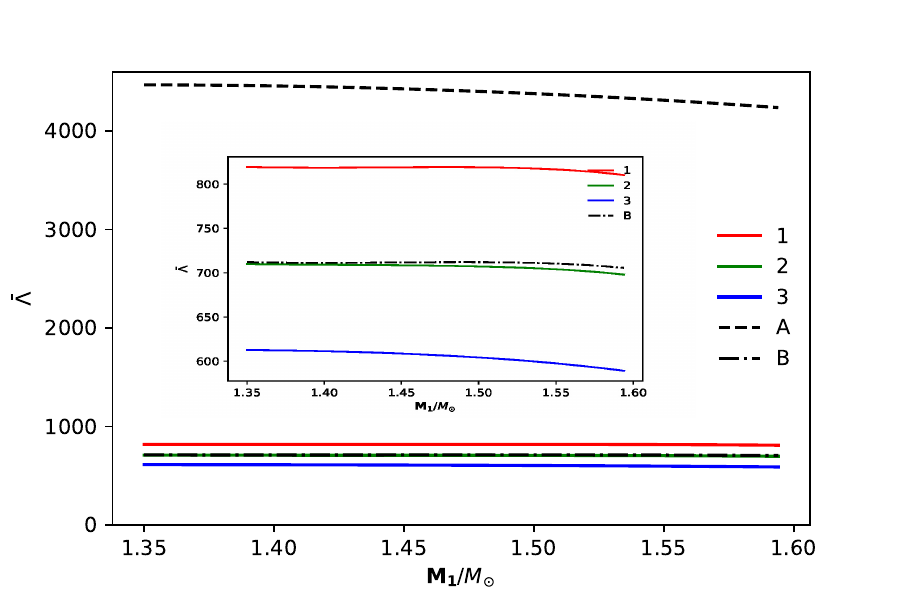}
    \caption{Tidal deformability of different EoS ($\bar{\Lambda}$) vs one of the
masses $\rm{M}_1$) for partial sets of the values for $\ell^{-7}$ used in FIG. \ref{fig1}.}
    \label{fig3}
\end{figure}

The boundary of tidal deformability from GW170817 and the properties of the five EoS will be stated more detailed now. 
FIG. \ref{fig2} shows the $\Lambda$-M relation for these different NS. The curves are plotted by combining with EoS and Eqs.(\ref{tln}). Detailed derivation of tidal deformability is referred to Appendix B.  As introduced above, the analysis of GW170817 impose a constraint of tidal deformability. For the NS binary system in this event, we can get the combined deformability with the two mass $M_{1,2}$ ($M_1 > M_2$) and their respective tidal deformability $\Lambda_{1,2}$
\ba\label{36}
{\bar{\Lambda}}=\frac{16}{13} \frac{\left(M_1+12 M_2\right) M_1^4 \Lambda_1+\left(M_2+12 M_1\right) M_2^4 \Lambda_2}{\left(M_1+M_2\right)^5}.
\ea
In \cite{LIGOScientific:2018hze}, a window $\bar{\Lambda}=300^{+420}_{-230}$ is given. So we choose three values for instanton gas model around $\bar{\Lambda}=700$ as shown in FIG. \ref{fig3}.  From the analysis of event GW170817, we know that the lowest mass of two NS is $1.1 M_\odot$ while the total mass is $2.7 M_\odot$. So the mass ranges from $1.35 M_\odot$ to $1.6 M_\odot$ in FIG. \ref{fig3}.  Compared among all the three instanton gas EoS, we found that the green line is the allowed realistic model with maximum mass, because the curve satisfies the upper bound of $\bar{\Lambda}$. Compared with curve 3, the maximal mass of curve 2 is larger, which is closer to the predicted ceiling of NS mass. While for curve 1, though with higher maximum mass, its tidal deformability exceeds the constraint. For further comparison, we set the same $\ell^{-7}$ for curve $\textit{A}$. It can be seen that the EoS violates the windows much from all the three figures. We also adjust $\bar{\Lambda}$ of curve $\textit{B}$ to the similar with curve 2, with the value $\ell^{-7}=17000$, as shown in the pointlike case \cite{Zhang:2019tqd}. All those figures imply that, the instanton gas model indeed shows better results,reproducing a maximum $1.85 M_\odot$, while satisfying the GW constraints on tidal deformability. 

\section{IV. Conclusions}
The WSS model derived from AdS/QCD offers an holographic way to deal with high density nuclear matter, which is difficult in traditional approach. In this letter, we apply the instanton gas structure in WSS model to extract the relation between the neutron energy density and the pressure, i.e., the EoS. This is an update for the point-like case in \cite{Zhang:2019tqd}, with the instanton width $\rho$ as an extra variable. One thing to emphasize is that, $\rho$ is found to be also a function of the chemical potential $\mu$, rather than a free parameter. Thus, the EoS obtained still has only one adjustable parameter, the D8- and  $\overline{\rm D8}$-branes separation distance $\ell$.
By applying the constraint from GW observation data, especially the binary NS event GW170817 (other NS events offer less sharp windows for TLN), we find that the maximum support mass is about $1.85 M_\odot$. This value is below the known criterion of known observation of $2 M_\odot$ NS, but still offer a brand new direction to finally solve the theoretical derivation of EoS in future.

\noindent {\it Acknowledgements.} 
The authors thank Prof. Feng-Li Lin and Zhoujian Cao  for helpful discussions.  KZ (Hong Zhang) is supported by a classified fund from Shanghai city.

\appendix
\section{A. Numerical Calculation Setup}\label{ap1}

In practice, the process of numerical calculation is constructed as follows. To obtain the pressure $p$ and energy density $\epsilon$ in the EoS, we need to solve the following three equations.

\begin{equation}\label{V16}
0 = \int_{0}^{\infty} \frac{2z I_1(z, n_I, \rho, k)}{3(1 + z^2)^{\frac{2}{3}}} \, dz,
\end{equation}

\begin{equation}
0 = \int_{0}^{\infty} \frac{2z I_2(z, n_I, \rho, k)}{3(1 + z^2)^{\frac{2}{3}}} \, dz,
\end{equation}

\begin{equation}
\mu(n_I, \rho, k) = \int_{0}^{\infty} \frac{2z I_4(z, n_I, \rho, k)}{3(1 + z^2)^{\frac{2}{3}}} \, dz.
\end{equation}
Then we can read off
\begin{equation}\label{pre}
p(n_I, \rho, k) = \frac{2}{7} - \int_{0}^{\infty} \frac{2z I_5(z, n_I, \rho, k)}{3(1 + z^2)^{\frac{2}{3}}} \, dz + \mu(n_I, \rho, k) \cdot n_I,
\end{equation}
and
\begin{equation}\label{eps}
\epsilon(n_I, \rho, k) = -\frac{2}{7} + \int_{0}^{\infty} \frac{2z I_5(z, n_I, \rho, k)}{3(1 + z^2)^{\frac{2}{3}}} \, dz.
\end{equation}

In which, for the convenience of numerical integration in the equations, we can change the variable $u$ to $z$ with $z$ defined as $u(z) = (1 + z^2)^{\frac{1}{3}}$ before proceeding with the numerical integration. 

In Eq. \eqref{V16}-\eqref{eps}, the functions $I_i$ are respectively

\begin{equation}
    \begin{split}\label{V20}
I_1(z, n_I, \rho, k) =& \left(\frac{3 n_I \rho^3 (-4 + 4 u(z)^3 - \rho^2)}{8 u(z)^{\frac{1}{2}} (-1 + u(z)^3 + \rho^2)^{\frac{7}{2}}}\right) \\
&\times \left(\frac{u(z)^4}{f_c(z) \zeta
(z, n_I, \rho, k)} + \zeta
(z, n_I, \rho, k)\right)\\
&- \frac{3 n_I^2 f_c(z) \rho^3 u(z)^3 (-2 + 2 u(z)^3 + 3 \rho^2)}{4 u(z)^{\frac{5}{2}}(-1 + u(z)^3 + \rho^2)^4}\\
&\times \frac{\zeta
(z, n_I, \rho, k)}{4 u(z)^{\frac{5}{2}}(-1 + u(z)^3 + \rho^2)^4},
    \end{split}
\end{equation}

\begin{equation}
    \begin{split}
I_2(z, n_I, \rho, k) = &
\frac{3 n_I \rho^4}{16 u(z)^{\frac{1}{2}} (-1 + u(z)^3 + \rho^2)^{\frac{7}{2}}} \\
&\times ((7 + 8 u(z)^3 + 3 \rho^2) \zeta
(z, n_I, \rho, k)\\
&- \frac{u(z)^7 (21 - 11 \rho^2 - 21 u(z)^3 + 5 u(z)^3 \rho^2)}{(-1 + u(z)^3)^2 \zeta
(z, n_I, \rho, k)}) \\
&- \frac{3 n_I^2 \rho^4 (u(z)^3 + 2) (-2 + 2 u(z)^3 + 3 \rho^2)}{8 u(z)^{\frac{5}{2}} (-1 + u(z)^3 + \rho^2)^4} \\
&\times \frac{\zeta(z, n_I, \rho, k)}{8 u(z)^{\frac{5}{2}} (-1 + u(z)^3 + \rho^2)^4} \\
&- \frac{n_I^{\frac{1}{2}} \left(1 + \frac{3 n_I}{4 \rho} - k^2\right)^{\frac{1}{2}}}{4 \rho^{\frac{1}{2}} (u(z) - 1)^{\frac{3}{2}}},
    \end{split}
\end{equation}

\begin{equation}
I_3(z, n_I, \rho, k) = \frac{k \zeta(z, n_I, \rho, k)}{u(z)^{\frac{11}{2}}},
\end{equation}

\begin{equation}
    \begin{split}
I_4(z, n_I, \rho, k) =&\left(\frac{3 \rho^4}{8 u(z)^{\frac{1}{2}} (-1 + u(z)^3 + \rho^2)^{\frac{5}{2}}}\right)\\
&\times \left(\frac{u(z)^7}{(u(z)^3 - 1) \zeta
(z, n_I, \rho, k)} + \zeta
(z, n_I, \rho, k)\right) \\
&+ \frac{n_I (u(z)^3 - 1) (-2 + 2 u(z)^3 + 3 \rho^2)^2}{4 u(z)^{\frac{5}{2}} (-1 + u(z)^3 + \rho^2)^3} \\
&\times \frac{\zeta
(z, n_I, \rho, k)}{4 u(z)^{\frac{5}{2}} (-1 + u(z)^3 + \rho^2)^3},
    \end{split}
\end{equation}
and
\begin{equation}
    \begin{split}\label{V24}
I_5(z, n_I, \rho, k) =& - u(z)^{\frac{5}{2}}+u(z)^{\frac{5}{2}} \cdot \zeta
(z, n_I, \rho, k) \\
&\times \left(1 + g_2(z, n_I, \rho) + \frac{(n_I \cdot Q(z, \rho))^2}{u(z)^5}\right).
    \end{split}
\end{equation}

The $- u(z)^{\frac{5}{2}}$ term in $I_5$, together with the $\frac{2}{7}$ terms in \eqref{pre} \eqref{eps}, are important, which are introduced to cancel the divergences at infinity. Those do not affect the physical meaning, since the divergent parts never change.

The terms $f_c(z)$, $Q(z, \rho)$, $q(z, \rho)$, $g_1(z, n_I, \rho)$, $g_2(z, n_I, \rho)$ and $\zeta(z, n_I, \rho, k)$ in Eq. \eqref{V20}-\eqref{V24} are defined as
\begin{equation}
f_c(z) = 1 - u(z)^{-3},
\end{equation}
\begin{equation}
Q(z, \rho) = \frac{(u(z)^{\frac{3}{2}}) (f_c(z)^{\frac{1}{2}}) (3\rho^2 + 2u(z)^3 - 2)}{2((-1 + u(z)^3 + \rho^2)^{\frac{3}{2}})},
\end{equation}
\begin{equation}
q(z, \rho) = \frac{9  \rho^4 
 u(z)^{\frac{1}{2}} }{4  f_c(z)^{\frac{1}{2}}  (-1 + u(z)^3 + \rho^2)^{\frac{5}{2}}},
\end{equation}
\begin{equation}
g_1(z, n_I, \rho) = \frac{u(z)^{\frac{1}{2}}  n_I  q(z, \rho)}{3  f_c(z)^{\frac{1}{2}}},
\end{equation}
\begin{equation}
g_2(z, n_I, \rho) = \frac{f_c(z)^{\frac{1}{2}}  n_I  q(z, \rho)}{3  u(z)^{\frac{7}{2}}},
\end{equation}
and
\begin{small}
\begin{equation}
\zeta(z, n_I, \rho, k) = \frac{\left(1 + g_1(z, n_I, \rho)\right)^{\frac{1}{2}}}{\left(1 + g_2(z, n_I, \rho) - \frac{k^2}{u(z)^8} + \frac{(n_I \cdot Q(z, \rho))^2}{u(z)^5}\right)^{\frac{1}{2}}}.
\end{equation}
\end{small}

\section{B. Tidal Love Number} \label{ap2}
 Under Boyer-Lindquist coordinate $(t,r,\theta.\phi)$ \cite{1967JMP.....8..265B}, we describe the metric of the sphericlal static NS with
 \begin{small}
\be
d s^2=g_{\alpha \beta}^{(0)} d x^\alpha d x^\beta=-e^{\nu(r)} d t^2+e^{\lambda(r)} d r^2+r^2\left(d \theta^2+\sin ^2 \theta d \phi^2\right) .
\ee
\end{small}
Setting $G_N=c=1$, $\nu(r)$ and $\lambda(r)$ in the above equation are obtained by $\nu(r)\equiv 2\phi(r)=-\lambda(r)=\ln(1-\frac{2 m(r)}{r})$.

The complete metric is combined with the two parts
\bea
g_{\alpha \beta}=g_{\alpha \beta}^{(0)}+h_{\alpha \beta},
\eea
where $h_{\alpha \beta}$ is a linearized metric perturbation.

Under Reege-Wheeler gauge, the perturbation has the expression \cite{Hinderer_2009}
\bea
h_{\alpha \beta}
=\operatorname{diag}[e^{-\nu(r)} H_0(r), e^{\lambda(r)} H_2(r),\nn\\
r^2 K(r), r^2 \sin ^2 \theta K(r)] Y_{2 m}(\theta, \varphi) .
\eea
Here $H_0(r)=H_2(r)=H(r)$ and $K(r)$ is also related to $H(r)$ by solving 
perturbative Einstein equation. On the other hand, $H(r)$ can be derived by solving the following equation
\begin{small}
\bea
\begin{aligned}
&H^{\prime \prime}+H^{\prime}\left[\frac{2}{r}+e^\lambda\left(\frac{2 m(r)}{r^2}+4 \pi r(p-\rho)\right)\right] 
\\
 &+H\left[-\frac{6 e^\lambda}{r^2}+4 \pi e^\lambda\left(5 \rho+9 p+\frac{\rho+p}{(d p / d \rho)}\right)-\nu^{\prime 2}\right]=0,
 \end{aligned}
\eea
\end{small}
At the core of NS where $r=0$, $H(r)$ has a solution
\bea
\begin{aligned}
H(r)=&\mathcal{B} r^2[1-\frac{2 \pi r^2}{7}(5 \rho(0)+9 p(0)+\frac{\rho(0)+p(0)}{(d p / d \rho)(0)}) \\
+&O(r^3)]\ ,
\end{aligned}
\eea
with $\mathcal{B}$ a constant and $dp/d\rho$ depends on the EoS.

At the surface of NS where $r$ is equal to the whole radius $R$, we get the Tidal Love Number
\begin{small}
\bea
\begin{aligned}
k_2\equiv &\frac{8 C^5}{5}(1-2 C)^2[2+2 C(y-1)-y]
\\
&\times\{4 C^3[13-11 y+C(3 y-2)+2 C^2(1+y)]
\\
&+3(1-2 C)^2 
[2-y+2 C(y-1)] \log (1-2 C)
\\
&+2 C(6-3 y+3 C(5 y-8))\}^{-1} .
\end{aligned}
\eea
\end{small}
Here the compactness of star $C \equiv M / R$ with the total mass $M$. The quantity $y \equiv R H^{\prime}(R) / H(R)$ is set for simplicity in calculation.  Actually, we can alternatively  start from the evaluation of $Y(r)\equiv  r H^{\prime}(r) / H(r)$ directly, then the second order differential equation becomes first order, and easier to solve, as shown in \cite{Postnikov:2010yn}.

The dimensionless tidal deformability $\Lambda$ is is related to the $l=2$ tidal Love number $k_2$ by
\be
\Lambda=\frac{2}{3} k_2\left(\frac{R}{M}\right)^{5} \text {. }
\ee

\section{C. Parameter Sensitivity Analysis} \label{ap3}
For the equation of state \eqref{eos1} in the instanton gas model, we can write it as follows
{\small
\bea
{\epsilon_1 \over \epsilon_{\odot}}=
\kappa_1 \mathcal{A}^{0.571}\left({p_1 \over p_{\odot}}\right)^{\gamma_1}+\kappa_2 \mathcal{A}^{-0.335}\left({p_1 \over p_{\odot}}\right)^{\gamma_2} .
\eea
}

 To analyze the sensitivity of the parameters, 
We change the four parameters $\kappa_1=0.140$, $\gamma_1=0.429$, $\kappa_2=3.896$ and $\gamma
_2=1.335$ by $\pm 5\%$ respectively.
 The sensitivity of each parameter can be checked by comparing the changes of the results, and the key parameters are identified.

Specifically, we take the Green line of the result curve 2 obtained as the comparison line, and then do the same treatment for each of the four parameters, that is, 
first increase the parameter value by $5\%$ as shown in the Orangered curve $i$, and then decrease the parameter value by $5\%$ as shown in the Deepskyblue curve $j$.

\begin{figure}[htbp!]
\centering
\subfigure[ Mass ($\rm{M}$) vs Radius ($\rm{R}$) relations for different $\kappa_1$]
{\begin{minipage}[t]{0.86\linewidth}
\centering
\includegraphics[width=0.8 \textwidth]{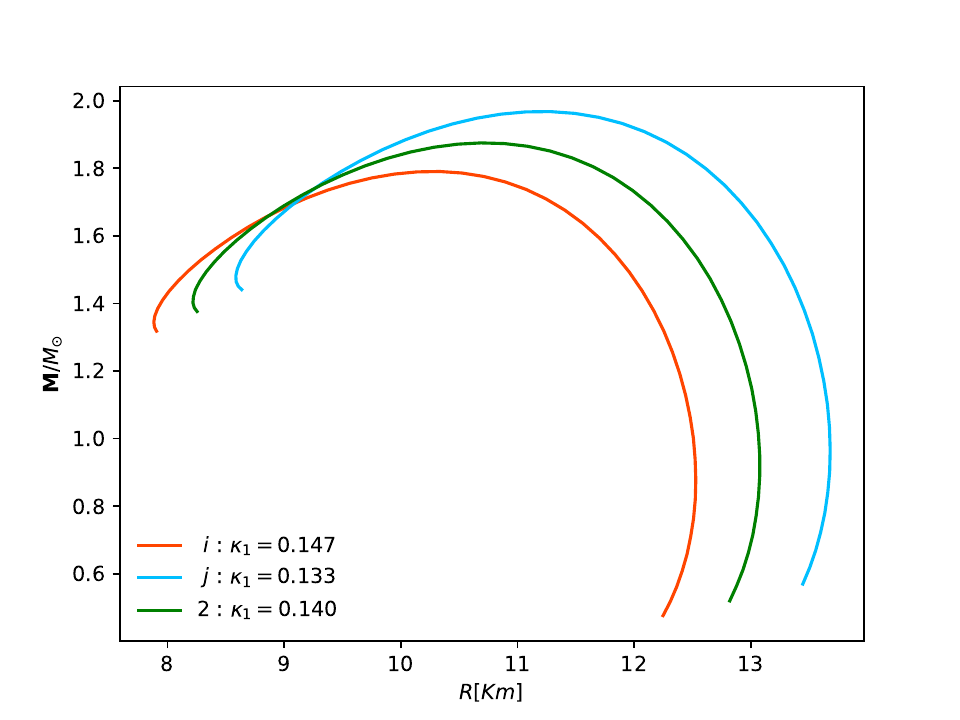}
\end{minipage}
}
\subfigure[Tidal deformability ($\Lambda$) vs Mass ($\rm{M}$) relations for different $\kappa_1$]
{\begin{minipage}[t]{0.86\linewidth}
\centering
\includegraphics[width=0.8\textwidth]{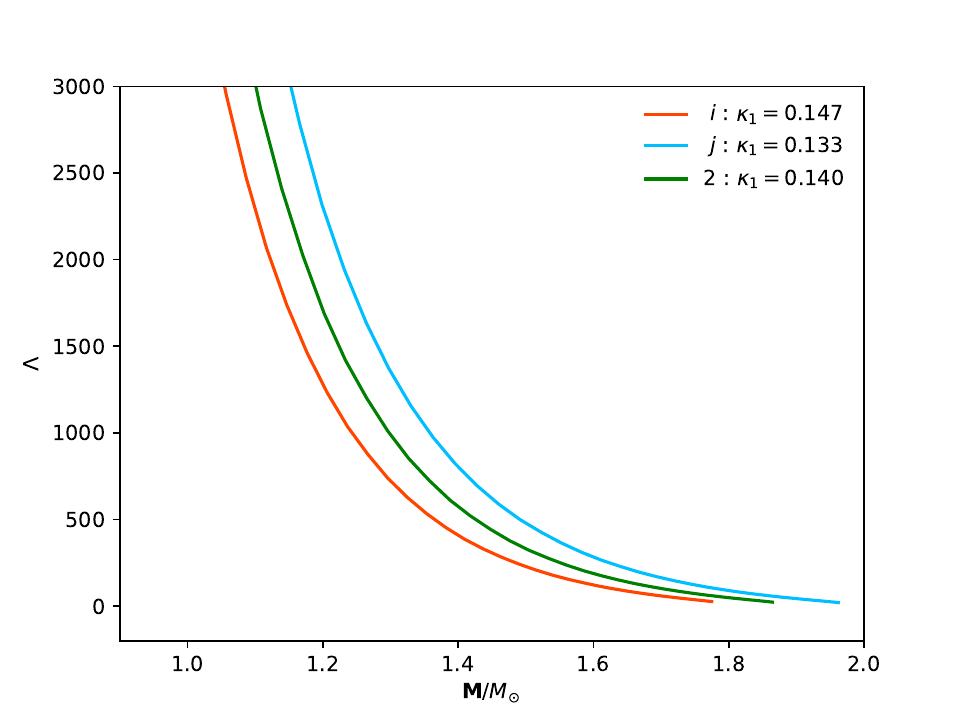}
\end{minipage}
}
\caption{For different $\kappa_1$ in EoS \eqref{eos1} }\label{diffk1}
\end{figure}

\begin{figure}[htbp!]
\centering
\subfigure[Mass ($\rm{M}$) vs Radius ($\rm{R}$) relations for different $\gamma_1$]
{\begin{minipage}[t]{0.86\linewidth}
\centering
\includegraphics[width=0.8 \textwidth]{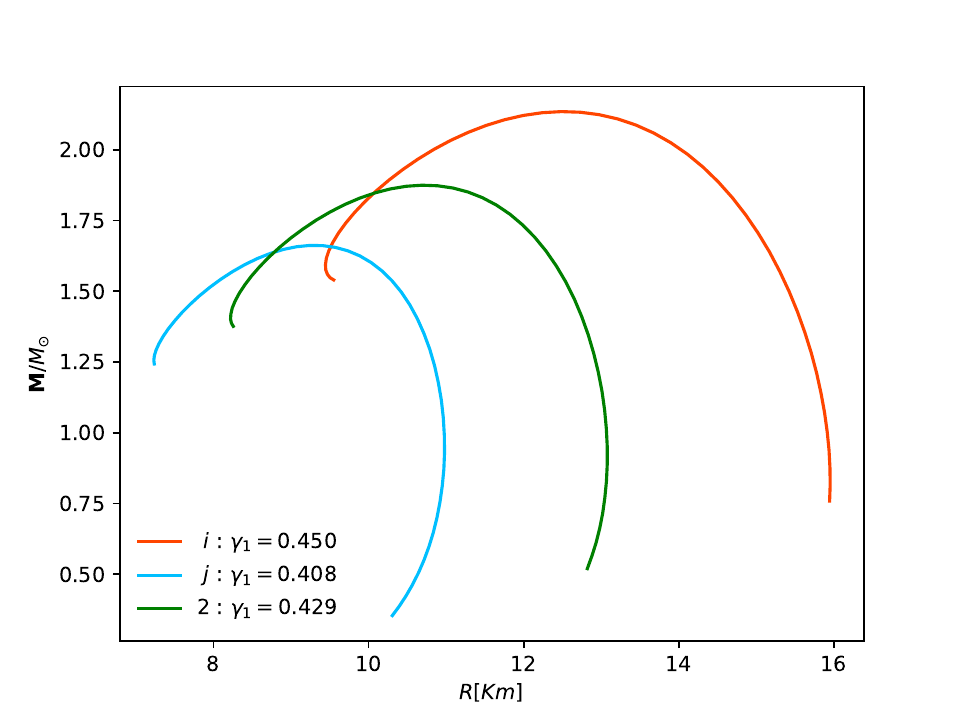}
\end{minipage}
}
\subfigure[Tidal deformability ($\Lambda$) vs Mass ($\rm{M}$) relations for different $\gamma_1$]
{\begin{minipage}[t]{0.86\linewidth}
\centering
\includegraphics[width=0.8\textwidth]{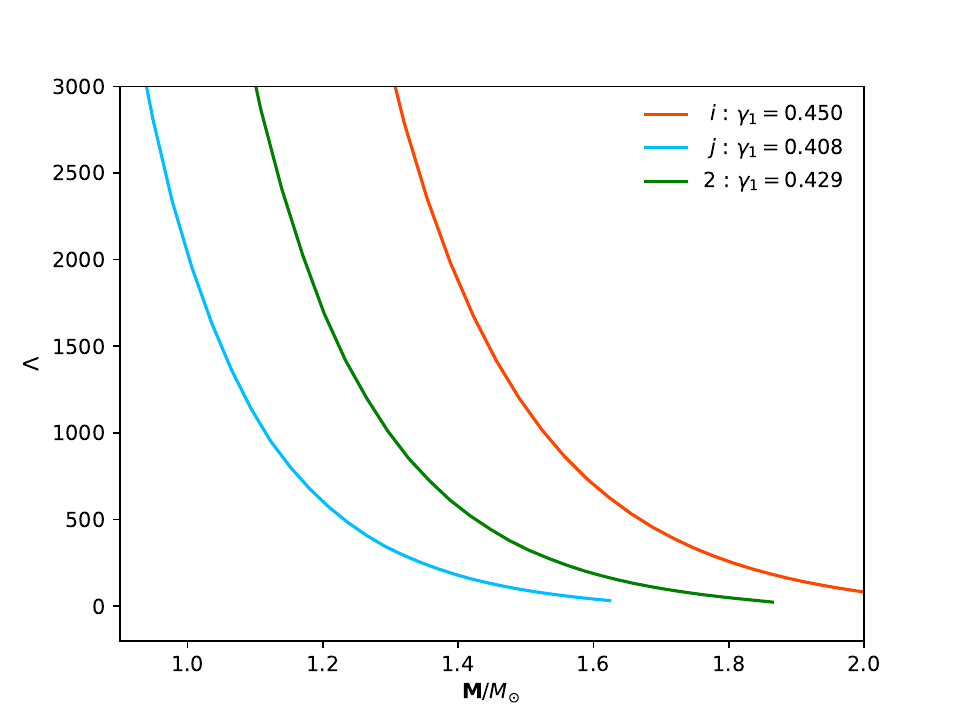}
\end{minipage}
}
\caption{For different $\gamma_1$ in EoS \eqref{eos1}}\label{diffg1}
\end{figure}

\begin{figure}[htbp!]
\centering
\subfigure[Mass ($\rm{M}$) vs Radius ($\rm{R}$) relations for different $\kappa_2$]
{\begin{minipage}[t]{0.86\linewidth}
\centering
\includegraphics[width=0.8 \textwidth]{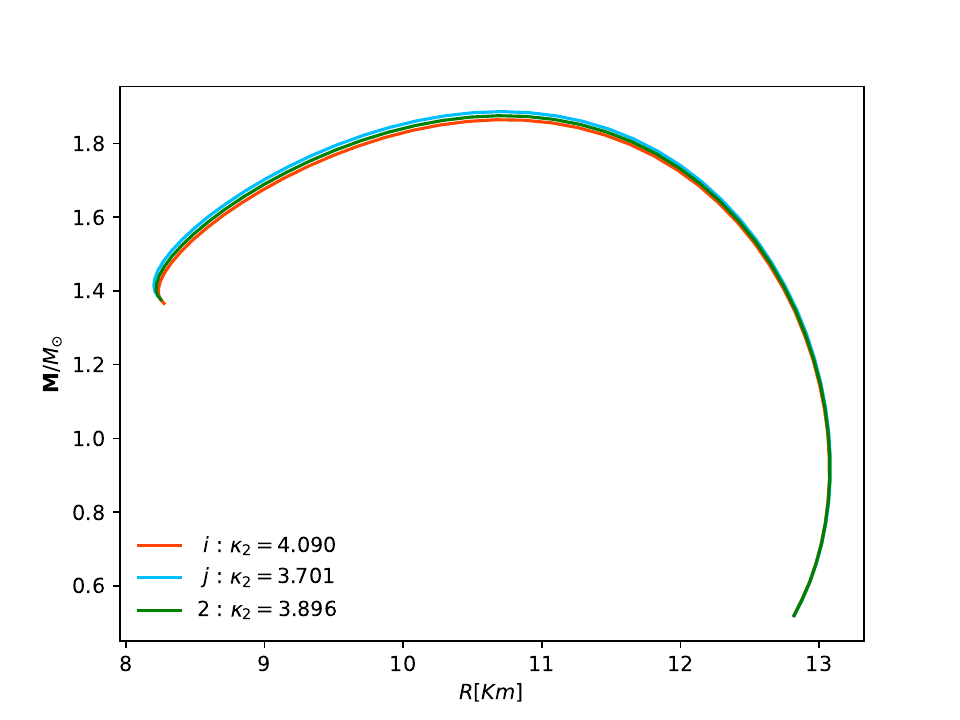}
\end{minipage}
}
\subfigure[Tidal deformability ($\Lambda$) vs Mass ($\rm{M}$) relations for different $\kappa_2$]
{\begin{minipage}[t]{0.86\linewidth}
\centering
\includegraphics[width=0.8\textwidth]{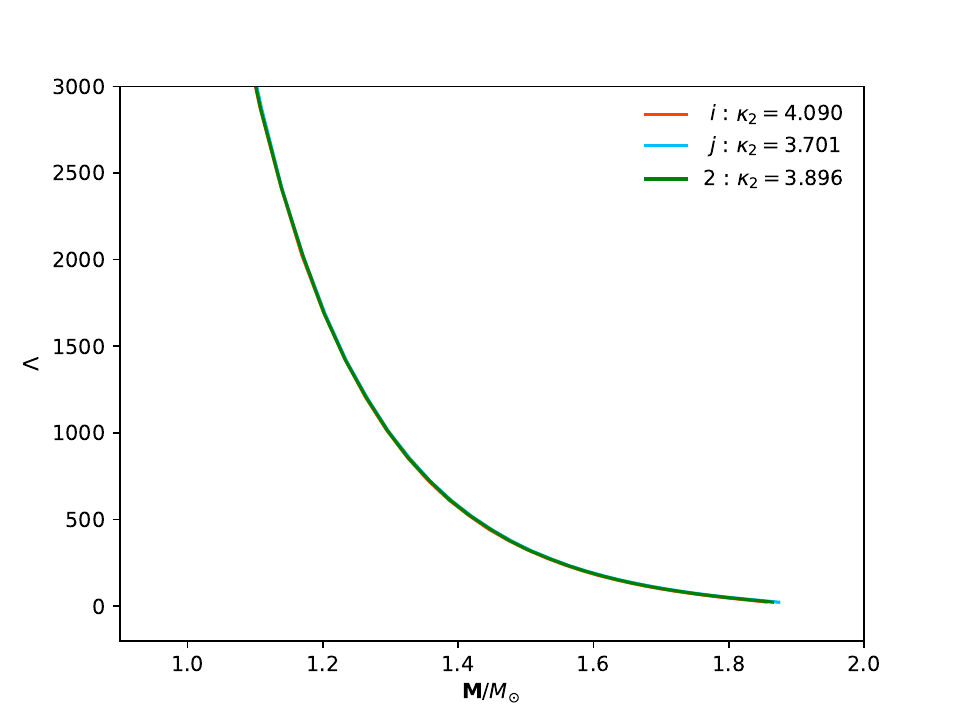}
\end{minipage}
}
\caption{For different $\kappa_2$ in EoS \eqref{eos1}}\label{diffk2}
\end{figure}

\begin{figure}[htbp!]
\centering
\subfigure[Mass ($\rm{M}$) vs Radius ($\rm{R}$) relations for different $\gamma_2$]
{\begin{minipage}[t]{0.86\linewidth}
\centering
\includegraphics[width=0.8 \textwidth]{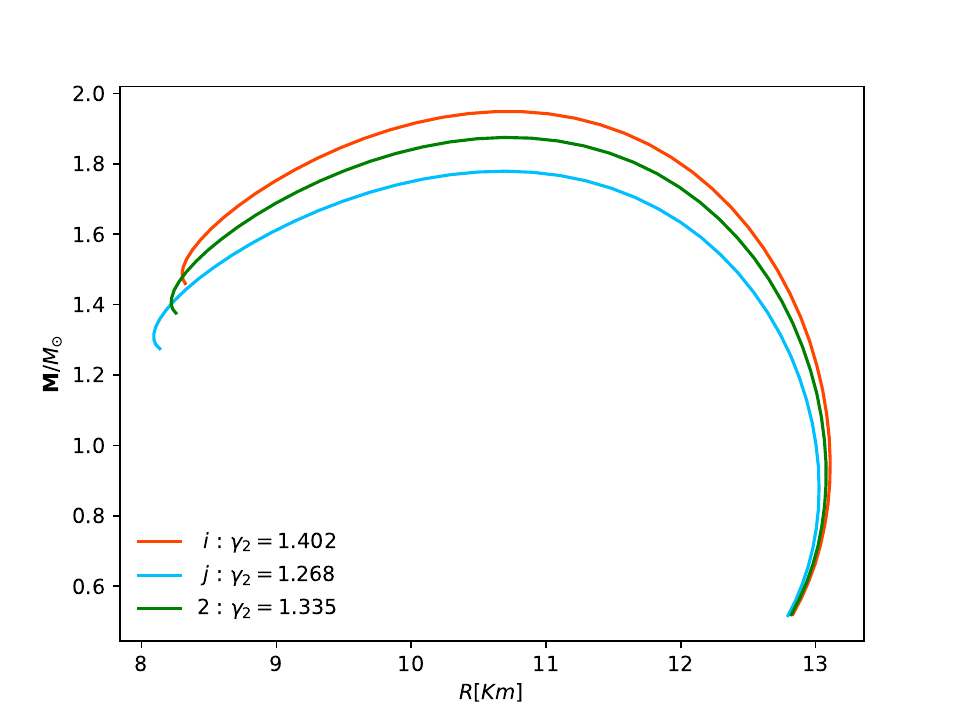}
\end{minipage}
}
\subfigure[Tidal deformability ($\Lambda$) vs Mass ($\rm{M}$) relations for different $\gamma_2$]
{\begin{minipage}[t]{0.86\linewidth}
\centering
\includegraphics[width=0.8\textwidth]{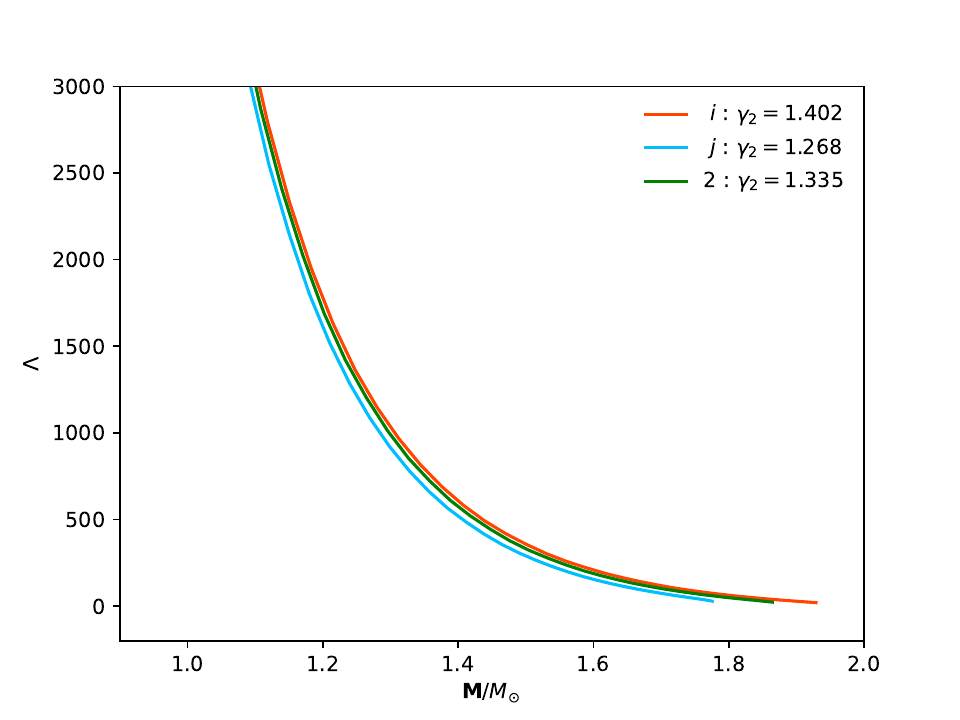}
\end{minipage}
}
\caption{For different $\gamma_2$ in EoS \eqref{eos1}}\label{diffg2}
\end{figure}

After comparison,
 we find that the parameter $\kappa_2$ has minor effect on the $\rm{M}$-$\rm{R}$ relationship and the $\Lambda$-$\rm{M}$ relationship as shown in Figure.\ref{diffk2}.
Parameters $\kappa_1$ and $\gamma_1$ have much greater influence on the $\rm{M}$-$\rm{R}$ relationship and  $\Lambda$-$\rm{M}$ relationship.
The change of parameter $\kappa_1$ is negatively correlated with the change of $\rm{M}$-$\rm{R}$ relation and $\Lambda$-$\rm{M}$ relation as shown in Figure.\ref{diffk1}, 
while in Figure.\ref{diffg1} the change of parameter $\gamma_1$ is positively correlated with the change of $\rm{M}$-$\rm{R}$ relationship and  $\Lambda$-$\rm{M}$ relationship. 
For the change of parameter $\gamma_2$, the situation is more special as illustrated in the Figure.\ref{diffg2}. The change of $\gamma_2$ has an effect on the $\rm{M}$-$\rm{R}$ relationship, but has a very small influence on the $\Lambda$-$\rm{M}$ relationship.

When the parameter $\kappa_1$ is set to 0, the $\rm{M}$-$\rm{R}$ relationship and the $\Lambda$-$\rm{M}$ relationship cannot be plotted, since in this case there is no solution to the TOV equations. 

\begin{figure}[htbp!]
\centering
\subfigure[Mass ($\rm{M}$) vs Radius ($\rm{R}$) relations]
{\begin{minipage}[t]{0.86\linewidth}
\centering
\includegraphics[width=0.8 \textwidth]{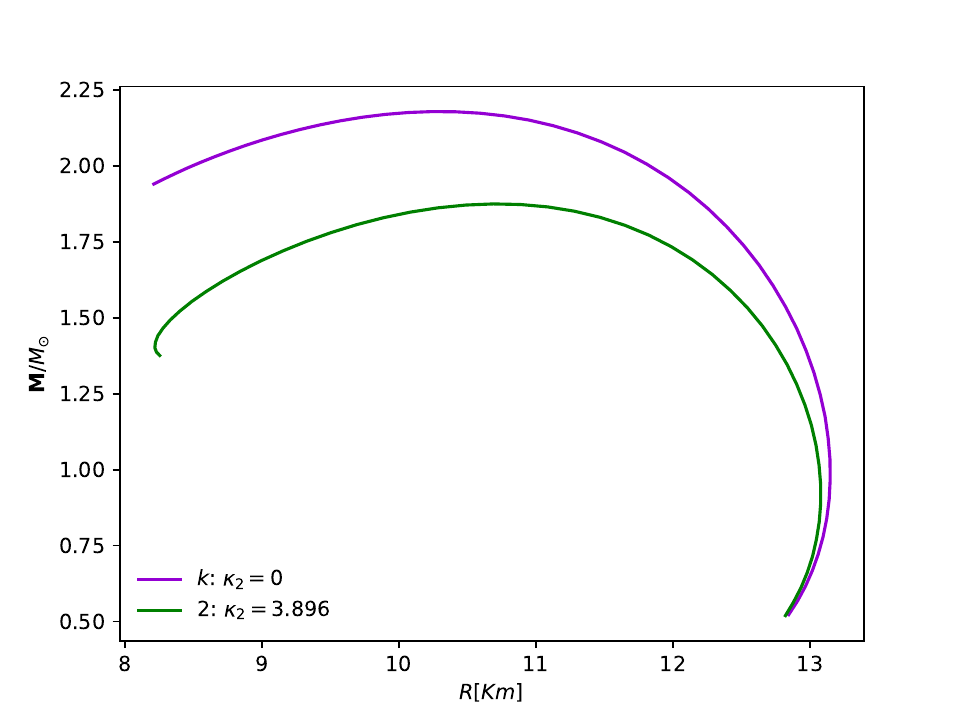}
\end{minipage}
}
\subfigure[Tidal deformability ($\Lambda$) vs Mass ($\rm{M}$) relations]
{\begin{minipage}[t]{0.86\linewidth}
\centering
\includegraphics[width=0.8\textwidth]{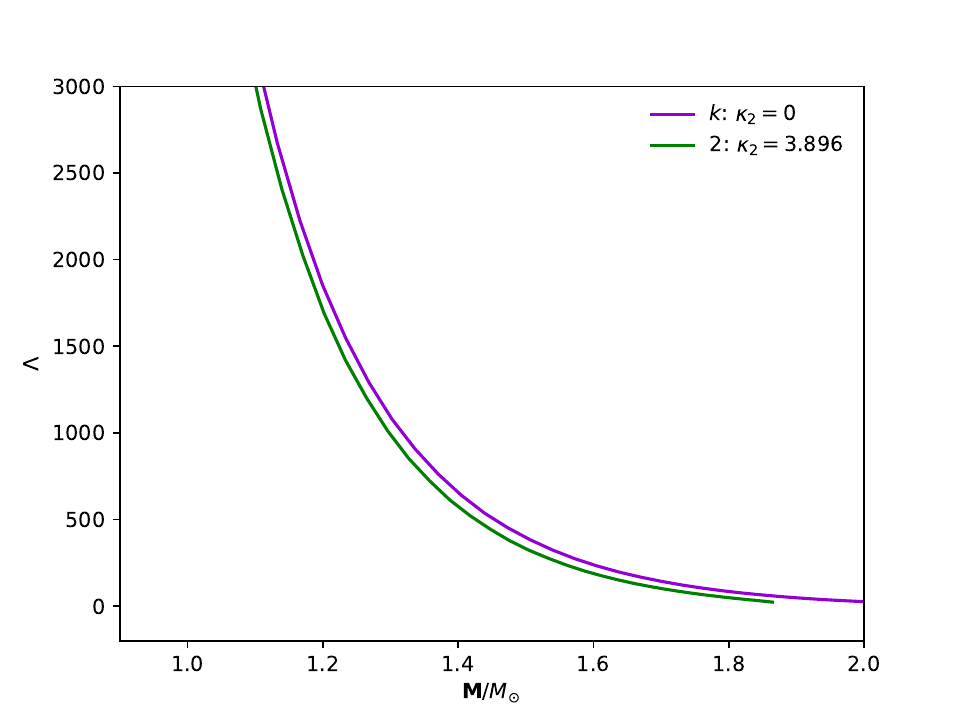}
\end{minipage}
}
\caption{Compared results for special vlaues of $\kappa_2$ in EoS \eqref{eos1}.}\label{k2is0}
\end{figure}

When the parameter $\kappa_2$ is 0, the corresponding curve is represented by the Darkviolet curve $k$. The results in Figure.\ref{k2is0} show that $\kappa_2 = 0$ has an effect on the $\rm{M}$-$\rm{R}$ relationship, but the effect on the $\Lambda$-$\rm{M}$ relationship is small.

in sum, we find that for the equation of state \eqref{eos1} in the instanton gas model, parameters $\kappa_1$ and $\gamma_1$ are the more critical factors.

\subsection{C.1 Scaling Symmetry}
We mention that there is a scaling symmetry \cite{Maselli:2017vfi, Zhang:2023hxd} in mass-radius and TLN relation. For a double-polytropic EoS $\epsilon(p) = \kappa_1 p^{\gamma_1} + \kappa_2 p^{\gamma_2}$, when we multiply the parameters $\kappa_1$ and $\kappa_2$
by a factor of $\mathcal{B}$ (e.g. $\ell^{-7}$), the energy density turns as $\epsilon \to \mathcal{B}\epsilon$.
Then from TOV equations (\ref{tov}), we see the quantities rescale with the factor $\mathcal{B}$ as $p \to \mathcal{B}p$, $m \to m / \sqrt{\mathcal{B}}$ and $r \to r / \sqrt{\mathcal{B}}$.
But TLN is invariant under the rescaling, since it is a function of purely $C$ and $y$, which are unaffected by the rescaling, as can be seen from their definition in Appendix. B.
With scaling symmetry, we can change the mass to our expected value more conveniently.

\bibliographystyle{unsrt}
\bibliography{Instanton_gas}

\end{document}